
%
\input phyzzx
%
%
\def\frac#1#2{{{ #1 \over #2 }}}      
%
%
\def\D{{\cal D}}
\def\OO{{\cal O}}
%
%
\def\mn{{\mu\nu}}
\def\a{\alpha}
\def\b{\beta}
\def\d{\delta}
\def\ga{\gamma}
\def\m{\mu}
\def\n{\nu}
\def\r{\rho}
\def\s{\sigma}

\def\dd{\Delta}

\def\ee{\epsilon}

\def\gm{\Gamma}
\def\vm{v_\m}
\def\bvm{\bar{v}_\m}
%
%
\def\tp{\tilde{p}}
\def\hp{\hat{p}}
\def\VV{\scriptscriptstyle V}
\def\cv{c_{\VV}}
\def\RR{{\rm I\!\!\, R}}
\def\idx{\int d^3\!x\>}
\def\iddx{\int d^D\!x}
%
%
\def\A{A_{\m}^a}
%

\mathsurround=2pt
\Pubnum={NIKHEF-H 93-05 (To appear in Phys. Rev. D)}
\date{}


\REF\Witten{E. Witten, Comm. Math. Phys. {\bf 121} (1989) 351.}
\REF\Guada{E. Guadagnini, M. Martellini and M. Mintchev, Nucl.
Phys. {\bf 330B} (1990) 557.
\nextline
P. Cotta-Ramusino, E. Guadagnini, M. Martellini and M. Mintchev,
Nucl. {\bf 330B} (1990) 577.}
\REF\Dunne{For a canonical analysis in which quantization is prior
to reduction see G.V. Dunne, R. Jackiw and C.A. Trugenberger, Ann.
Phys. {\bf 194} (1989) 197.}
\REF\Birmingham{D. Birmingham, M. Rakowsky and G. Thompson,
Nucl. Phys. {\bf B329} (1990) 83.}
\REF\Gieres{F. Delduc, F. Gieres and S.P. Sorella, Phys. Lett.
{\bf B225} (1989)
367.}
\REF\Delduc{F. Delduc, C. Lucchesi, O. Piguet and S.P. Sorella,
Nucl. Phys. {\bf B346} (1990) 313.}
\REF\Blasi{A. Blasi and R. Collina, Nucl. Phys. {\bf B345} (1990)
472.}
\REF\Carmelo{C.P. Martin, Phys. Lett. {\bf B241} (1990)513.}
\REF\gmr{G. Giavarini, C.P. Martin and F. Ruiz Ruiz, Nucl. Phys.
{\bf 381B} (1992) 222.}
\REF\AGaume{L. Alvarez-Gaum\'e, J.M.F. Labastida and A.V. Ramallo,
Nucl. Phys. {\bf 334B} (1990) 103.}
\REF\Asorey{M. Asorey and F. Falceto, Phys. Lett. {\bf 241} (1990)
31.}
\REF\Guadareg{E. Guadagnini, M. Martellini and M. Mintchev, Phys.
Lett. {\bf B227} (1989) 111.}
\REF\Schonfeld{J. F. Schonfeld, Nucl. Phys. {\bf B185} (1981) 167.
\nextline
S. Deser, R. Jackiw and S. Templeton, Ann. Phys. {\bf 140} (1982)
372.}
\REF\Pisrao{R. D. Pisarski and S. Rao, Phys. Rev. {\bf D32}
(1985) 2081.}
\REF\Veltman{G. `t Hooft and M. Veltman, Nucl. Phys. {\bf B44} (1972)
189.}
\REF\Maison{P. Breitenlohner and D. Maison, Commun. Math. Phys.
{\bf 52} (1977) 11.}
\REF\Bonneau{G. Bonneau, Int. J. Mod. Phys. {\bf A5} (1990) 3831.}
\REF\BosWZW{M. Bos, Ann. Phys. {\bf 181} (1988) 177.}
\REF\Osborn{H. Osborn, Ann. Phys. {\bf 200} (1990) 1.}
\REF\Collins{J.C. Collins, {\it Renormalization}, Cambridge
University Press (1984).}
\REF\Bos{M. Bos and V.P. Nair, Mod. Phys. Lett. {\bf A5} (1990) 959.
\nextline
J.M.F. Labastida, P.M. Llatas and A.V. Ramallo, Nucl. Phys. {\bf B348}
(1991) 651.}
\REF\Speer{E.R. Speer, Ann. l'Inst. Henri Poincar\'e {\bf XXII}
(1975) 1.
\nextline
E.R. Speer, J. Math. Phys. {\bf 15} (1974) 1.}
\REF\reduce{REDUCE, The Rand Corporation, Santa Monica
(1991).}
\REF\Dorey{N. Dorey, Phys. Lett {\bf B246} (1990) 87.}
\REF\review{G. Giavarini, C.P. Martin and F. Ruiz Ruiz, {\it
Perturbative quantization of Chern-Simons theory}, preprint.}
\REF\Jaffe{A. Jaffe and A. Lesniewski, {\it Supersymmetric quantum
fields and infinite dimensional analysis}, in {\it Non-perturbative
quantum field theory}, G. 't Hooft, A. Jaffe, G. Mack, P.K. Mitter
and R. Stora eds., Plenum Press (1988).}
\REF\Epstein{H. Epstein and V. Glaser, Ann. Inst. Henri Poincar\'e
{\bf XIX} (1973) 211.}


\titlepage

\title{{\seventeenbf Physically meaningful and not so meaningful
symmetries in Chern-Simons theory}}

{\author{ G. Giavarini$^\star$}}\footnote{\hbox{$\star$}}{Address
after May 1, 1993: {\it INFN Gruppo collegato di Parma and Dipartimento
di Fisica dell'Uni\-ver\-sit\`a di Parma, Viale delle Scienze,
I-43100 Parma, Italy}}

\address{{\it Laboratoire de Physique Th\'eorique et Hautes Energies,
              Universit\'es Paris VII \break
              Tour 14-24, $5^{eme}$ \'etage,
              2 Place Jussieu, 75251 Paris Cedex 05, France}}

\author{ C. P.~ Martin}
\address{{\it Department of Mathematics and Statistics,
              University of Guelph \break
              Guelph, Ontario N1G 2W1, Canada}}

\author{ F.~ Ruiz Ruiz}
\address{{\it The Niels Bohr Institute, Blegdamsvej 17, DK-2100
              Copenhagen \O, Denmark \break
              and \break
              NIKHEF-H, Postbus 41882, 1009 DB Amsterdam,
              The Netehrlands$^{\star\star}$}}\footnote{
              \hbox{$\star\star$}}{Present address}

\vskip .7 true cm

\noindent
We explicitly show that the Landau gauge supersymmetry of
Chern-Simons theory does not have any physical significance.
In fact, the difference between an effective action both
BRS invariant and Landau supersymmetric and an effective
action only BRS invariant is a finite field redefinition.
Having established this, we use a BRS invariant regulator
that defines CS theory as the
large mass limit of topologically massive Yang-Mills theory to
discuss the shift $k \to k+\cv$ of the bare Chern-Simons parameter
$k$ in conncection with the Landau supersymmetry.
Finally, to convince ourselves that the shift above
is not an accident of our regularization method, we comment
on the fact that all BRS invariant regulators used as yet yield
the same value for the shift.

\endpage

\pagenumber=2

{\bf\chapter{Introduction}}

Canonical quantization of three-dimensional Chern-Simons (CS)
theory has provided two very interesting results [\Witten]. One is
the relation between the vacuum expectation values of the Wilson
loops of the theory and the intrinsically three-dimensional
characterizations of knot and link invariants. The other
one is a framework to understand properties of two-dimensional
conformal theory. In both issues, two features of CS theory play a
major part: its finiteness and the shift of the bare CS parameter
$k$
$$
k\to k+ {\rm sign}\,(k)\> c_{\VV} ~ ,
\eqn\shift
$$
$\cv$ being the quadratic Casimir operator in the adjoint
representation of the gauge group. For a variety of reasons,
one would like to understand these two properties from
a perturbative point of view. Among such reasons, we mention
firstly the fact that perturbative quantization has led to
explicit integral representations of knot and link invariants of
the type of Gauss' integral for the linking number of two curves
[\Guada]. And secondly, that perturbative quantization controls
gauge invariance for the quantum theory through BRS invariance,
which in a sense corresponds to first quantizing and then
constraining, the opposite approach to what is usual in
canonical quantization of CS theory [\Dunne].

In perturbative quantization, the quantum theory is constructed
by demanding it to have certain symmetries. The problem of
determining the symmetries that characterize the quantum theory
thus becomes a fundamental issue. Classically, the theory has two
symmetries: topological invariance or invariance under changes of
the spacetime metric, and gauge invariance. Topological invariance
is trivially established, for both the classical action [see eq.
(2.2)] and the observables [see eq. (2.7)] are independent of any
metric. However, to later quantize the theory one fixes the gauge
and gauge fixing needs of a choice of metric so that the explicit
metric independence of the classical action is lost. This
does not spoil classical topological invariance, since the
spacetime metric only enters in a BRS exact term and BRS exact
terms have no observable meaning. Though, one is left with BRS
as the only manifest symmetry of the classical
gauge-fixed theory. Not quite! It happens that the
gauge-fixed classical action in the Landau gauge has a new
symmetry, the so called Landau gauge supersymmetry
[\Birmingham,\Gieres]. This new symmetry has been used in Ref.
[\Delduc] to prove perturbative finiteness to all orders (see Ref.
[\Blasi] for an alternative proof), but on the other hand is a
symmetry in only the Landau gauge. The purpose of this paper is
to study the relevance of this symmetry.

It will turn out that the Landau gauge supersymmetry has no
relevance and that, furthermore, it does not play any r\^ole in
the construction of the quantum theory. We will show this in Sect.
2. To actually compute the shift of the bare CS parameter $k$ within
the perturbative framewrok one has to use a regularization
prescription. It happens that all BRS invariant regulators
used so far [\Witten,\Carmelo-\Asorey] produce at one loop the same
shift as in eq. \shift. However, Landau supersymmetric regulators
[\Guadareg] do not. Unfortunately, there is no known regulator
preserving both BRS invariance and the Landau gauge supersymmetry
simultaneously. In Sect. 3 we analyze the Landau gauge
supersymmetry breaking for a particular BRS invariant regulator
[\Carmelo,\gmr], the only one which has produced as yet a check of
the shift in eq. \shift\ at two loops. Finally, Sect. 4 contains our
conclusions as well as a discussion of the existence of a unique
parametrization for quantum CS theory.

{\bf\chapter{BRS invariance, the Landau gauge supersymmetry and
finite renormalizations}}

The CS action in the Landau gauge for a $SU(N)$ gauge connection
$\A$ on $\RR^3$ reads in the fundamental representation:
$$
S = S_{CS} + S_{GF} \>,
\eqn\starting
$$
where $S_{CS}$ is the classical CS action
$$
S_{CS}=- {ik\over 4\pi} \!\idx  \ee^{\m\r\n} \left(
        {1\over 2}\,\, A^a_{\m}\partial_{\r}A^a_{\n} +
        {1 \over 3!}\,f^{abc}A^a_\m A^b_\r A^c_\n  \right)
\eqn\csaction
$$
and $S_{GF}$ is the Landau gauge fixing term
$$
S_{GF} = \idx \big[ - b^a\partial A^a
                     + \bar{c}^a \partial^\m (D_\m c)^a \big] \> ~.
\eqn\gfixing
$$
The parameter $k$ in eq. \csaction\ is the classical or bare CS
parameter. As usual, $b^a$ denotes the Lagrange multiplier imposing
the gauge condition $\partial A^a=0,$ $c^a$ and $\bar{c}^a$ are
Faddeev-Popov ghosts and $D_\m^{ac} = \d^{ac}\partial_\m
+ f^{abc}A_\m^b$ is the covariant derivative. The structure
constants $f^{abc}$ are completely antisymmetric and are normalized
so that $f^{acd}f^{bcd}=\cv\d^{ab}.$ We will keep $\cv$ in the
notation although for $SU(N)$ one has the simple expression $\cv=N.$
The action in eq. \starting\ is invariant under BRS transformations
$$
\eqalign{
s\A & = (D_\m c)^a  \cr
sc^a & = - \frac 12 \,f^{abc} c^b c^c  \cr} \qquad\qquad
\eqalign{
sb^a & = 0 \cr
s \bar c^a & = b^a \>. \cr}
\eqn\brstrans
$$
Note that the gauge fixing term introduces a
metric thus spoiling the metric independence of the CS classical
action $S_{CS}.$ Classical topological invariance is nevertheless
guaranteed by the BRS exactness of $S_{GF},$
$$
S_{GF} = - \idx s\,\big( \bar c^a \partial A^a \big) \>,
$$
and the fact that BRS exact quantities are unobservable, \ie\
unphysical.

In addition to BRS invariance, the action $S$ has the following two
symetries [\Birmingham,\Gieres]:
$$
\eqalign{
\vm A^a_\n &= {4\pi i\over k} \, \ee_{\m\n\r}\, \partial^\r c^a  \cr
\vm c^a & = 0   \cr} \qquad\qquad
\eqalign{
\vm b^a & = -(D_\m c)^a  \cr
\vm\bar c^a & =  A^a_\m  \cr}
\eqn\susy
$$
and
$$
\eqalign{
\bvm A^a_\n &= -{4\pi i\over k} \, \ee_{\m\n\r}\, \partial^\r \bar c^a  \cr
\bvm c^a & = A^a_\m  \cr} \qquad\qquad
\eqalign{
\bvm b^a & = \partial_\m \bar c^a  \cr
\bvm\bar c^a & = 0 \>. \cr}
\eqn\susybar
$$
These two sets of symmetries are indistinctively called Landau gauge
supersymmetry. It is important to notice that $S_{CS}$ and $S_{GF}$
are not separately invariant under $\vm$ nor under $\bvm$, but that
it is the whole gauge-fixed action $S$ that is invariant. Furthermore,
the Landau gauge supersymmetry is only an invariance of the gauge-fixed
classical action in the Landau gauge, never of the Wilson loops
(the observables of the theory). To see the latter, we recall
the definition of the Wilson loop for a closed curve $C:$
$$
W(C) = {\rm tr}\> P\> {\rm exp}\>
   \biggl\{ \oint_C A^a_\m T^a dx^\m \biggr\} \>,
\eqn\Wilson
$$
$T^a$ being the generators of the Lie algebra of the gauge group.
It is obvious that $W(C)$ is not invariant under $\vm$ nor under $\bvm$.

Here we want to study the significance of these symmetries
for the quantum theory. It is obvious that a quantum CS theory
without BRS invariance would not make any sense. On the contrary,
one expects the Landau gauge supersymmetry not to have much
relevance, despite the fact it was useful in proving perturbative
finiteness [\Delduc]. We expect the latter on the basis that
something that only holds in a particular gauge can not have much
significance. In the sequel we show that one can introduce at will
a breaking of the Landau gauge supersymmetry at the quantum level
by simply performing finite wave function renormalizations.

To discuss BRS invariance at the quantum level, we introduce the
standard external fields $J^{a\m}$ and $H^a$ coupled respectively
to the non-linear BRS transforms $s\A$ and $sc^a$ so that the
gauge-fixed classical action becomes
$$
\gm_0 = S_{CS} + S_{GF} + S_{EF} \>,
\eqn\treeaction
$$
where
$$
S_{EF} = \idx \left[ J^{a\m} (D_\m c)^a
                   - \frac 12 \,f^{abc}H^ac^bc^c\, \right] \>.
$$
It is well known that symmetries at the quantum level are governed by
their corresponding Ward identities so what we need are the Ward
identities for the BRS symmetry and the Landau gauge sypersymmetries.
The Ward identity for the BRS symmetry or BRS identity takes in our
notation the form
$$
\idx \biggl( {\d \gm \over\d A^a_\m}\,{\d \gm \over\d J^{a\m}}
           + {\d \gm \over\d H^a}\, {\d \gm \over\d c^a}
           + b^a\,{\d \gm \over\d \bar{c}^a} \biggr) \, = 0 \>,
\eqn\stora
$$
where $\gm$ is the effective action. In turn, the Ward identities
for the Landau gauge supersymmetries in eqs. \susy\ and \susybar\
read
$$
\idx \biggl[ {4\pi i\over k}\,\ee_{\m\n\r}\,(\partial^\n c^a)\,
               {\d \gm \over \d A^a_\r}
           + {4\pi i\over k}\,\ee_{\m\n\r}\, (\partial^\n J^{a\r})\,
               {\d \gm \over\d H^a}
           - A^a_\m\,{\d \gm \over\d \bar c^a}
           + {\d \gm \over\d b^a}\, {\d \gm \over\d J^{a\m}}
     \biggr] = 0
\eqn\wardsusy
$$
and
$$
\eqalign{
\idx \biggl[ { 4\pi i\over k}\,\ee_{\m\r\n} \,
             (J^{a\n}-\partial^\n \bar c^a) & \,{\d \gm \over\d A^a_\r}
           - A^a_\m\,{\d \gm \over\d c^a}
           - (\partial_\m \bar c^a)\, {\d \gm \over\d b^a}
           - H^a\, {\d \gm \over\d J^{a\m}}
     \biggr]  \cr
& = \idx \biggl( {4\pi i\over k}\,\ee_{\m\n\r}J^{a\n}\,\partial^\r b^a
               + J^{a\n} \partial_\m A^a_\n
               - H^a\,\partial_\m c^a
         \biggr) \>, \cr }
\eqn\wardsusybar
$$
respectively. One also wants the choice of gauge to be preserved by
quantization so that one supplements these equations above
with the Ward identity
$$
{\d \gm \over\d b^a} + \partial A^a=0 \>.
\eqn\gauge
$$
This equation, together with eq. \stora, implies that
$$
\partial^\m {\d \gm \over\d J^{a\m}}\,-\,
      {\d \gm \over\d {\bar c}^a}\, =\,0 \> .
\eqn\constraint
$$

The effective action $\gm$ is an integrated functional of mass
dimension three and ghost number zero that depends on the fields
$\A,\,\,b^a,\,\,c^a,\,\,\bar c^a,\,\,J^{a\m}$ and $H^a$ and that
has local and non-local contributions. In perturbation theory,
$\gm$ is given by a loop expansion
$$
\gm = \sum_{n=0}^{\infty} \gm_n \>,
$$
where the zero order contribution $\gm_0$ is the tree-level action
in eq. \treeaction\ and $\gm_n$ stands for the order $\hbar^n$
correction. We want to find the most general structure of its local
part compatible with eqs. \stora-\constraint. So let us analyze each
one of these equations. Eq. \constraint\ implies that $\gm$ depends
on the fields $J^{a\m}$ and $\bar c^a$ through the combination
$J^{a\m}-\partial^\m \bar c^a.$ Eq. \gauge\ on its own implies that
$\gm$ will be of the form $\gm = \bar\gm - \idx b\,\partial A,$
with $\bar\gm$ an integrated functional with the same mass dimension
and ghost number as $\gm$ but independent of $b^a.$

The analysis of the BRS identity eq. \stora\ is more involved. As a
first step, it requires showing that it has a solution, or in a more
familiar language, that there is no BRS anomaly. That this is the
case was proved in Ref. [\Blasi]. The local part of $\Gamma_1$ can
be actually constructed using the method of induction and solving
the corresponding linearized equation
$$
\Delta \gm_1 = 0 \>,
\eqn\stability
$$
where $\Delta$ is the Slavnov-Taylor operator
$$
\Delta = \idx \left( {\d\gm_0 \over \d A^a_\m}\,{\d\over \d J^{a\m}}
                    + {\d\gm_0 \over \d J^{a\m}}\,{\d\over \d\A}
                    + {\d\gm_0 \over \d c^a}\,{\d\over \d H^a}
                    + {\d\gm_0 \over \d H^a}\,{\d\over\d c^a}
                    + b^a\,{\d\gm_0 \over \d \bar c^a} \right) \>.
$$
This operator is the quantum generalization of the BRS classical
operator $s$ and, as the latter, is nilpotent: $\Delta^2=0$. We have
already said that $\gm_1$ contains local and non-local contributions.
A thorough study of eq. \stability\ shows, however, that local
contributions in $\gm_1$ decouple from non-local ones [\gmr]
and gives for the local part of $\gm_1$ the expression [\Blasi,\gmr]:
$$
\gm^{\rm local}_1 = \a \, S_{CS}
   + \Delta \idx \left[ \b ( J^{a\m} - \partial^\m \bar c^a ) A^a_\m
                        - \ga H^a c^a \right]  \>,
\eqn\counter
$$
where $\a,\>\b$ and $\ga$ are arbitrary coefficients of order
$\hbar.$ In what follows we will omit the superscript ``local''
from the notation. Putting together $\gm_0$ and $\gm_1$ we obtain
the effective action up to order $\hbar:$
$$
\eqalign{
\gm =\, & - {ik\over 4\pi} \idx \ee^{\m\r\n}
\left[ \, {1\over 2}\, (1+ \a + 2\,\b)\,A^a_\m \partial_\r A^a_\n
     + {1\over 3!}\,(1 + \a + 3\,\b) \,f^{abc}
                        A^a_\m\,A^b_\r\,A^c_\n \right] \cr
& + \idx \biggl\{ - b^a \partial A^a
                 + (J^{a\m}-\partial^\m \bar{c}^a)
                   \left[ (1 - \b + \ga)\, \partial_\m c^a
                        + (1 + \ga)\,f^{abc}A^b_\m c^c \, \right]
         \biggr\}\cr
& - \idx \frac 12 \,(1 + \ga)\, f^{abc}H^a c^b c^c  \> . \cr}
\eqn\effective
$$
Note that the theory is not finite by power counting so to make
explicit computations one has to use a regularization method. As
is well known, any regularization method will introduce ambiguities
in Green functions which are divergent by power counting, whereas
Green functions already convergent by power counting will remain
unambiguous. It happens that the only Green functions which
diverge by power counting, hence the only sources of ambiguities,
involve fewer than four fields. We have seen that their
generating functional at one loop is given by eq. (2.16).
To find explicit values for the coefficients $\a,\>\b$ and $\ga$
one may use a BRS invariant regularization method, with different
methods yielding in general different values. Recall that the
theory being finite, though not by power counting, implies that
the $\a,\>\b$ and $\ga$ are finite after whatever regulator one
decides to use is removed.

The structure of $\gm_1$ in eq. \counter\ shows that there are two
types of radiative corrections. We have on the one hand radiative
corrections labeled by $\b$ and $\ga;$ they correspond to the
cohomologically trivial term
$$
\Delta X \equiv
\Delta \idx \left[ \b ( J^{a\m} - \partial^\m \bar c^a ) A^a_\m
               - \ga H^a c^a \right]
\eqn\brsexact
$$
and, hence, do not contribute to the vacuum expectation values
of the observables. On the other
hand, we have the radiative corrections labeled by $\a;$ they
correspond to the gauge invariant quantity $\a\,S_{CS}$ and
contribute to the vacuum expectation values of the
observables. The fact that radiative corrections
of the first type have the cohomologically trivial form $\Delta X$
ensures that they can be set to zero by renormalizing only the
fields. Indeed; any wave function renormalization of the form
$$
\Phi = Z_\Phi\,\Phi' \> ,
\eqn\schemeone
$$
with
$$
Z_A = Z^{-1}_b = 1 - \b \qquad
Z_c Z_{\bar c} = 1 + \b - \ga \qquad
Z_HZ_c^2 = 1 - \ga \qquad
Z_J = Z_{\bar c} \> ,
$$
absorbs the contribution $\Delta X$ to the effective action so
$\gm$ in terms of the renormalized fields $\Phi'$ writes
$$
\eqalign{
\gm'= & - {ik\,(1+\a) \over 4\pi} \idx \ee^{\m\r\n}
      \left( {1\over 2} \,A'^a_\m \partial_\r A'^a_\n
           + {1\over{3!}}\,f^{abc} A'^a_\m\,A'^b_\r\,A'^c_\n
      \right) \cr
& + \idx \left[ \,-\, b'^a \partial A'^a
                + (J'^{a\m}-\partial^\m \bar{c}'^a)\, (D'_\m c')^b
                - {1\over 2}\,f^{abc}H'^a c'^b c'^c \,
         \right] \>, \cr}
\eqn\actionone
$$
or more simply
$$
\gm' = \gm_0\, [\,\Phi'\,, \,k+\a\,] \>.
$$
We denote by ${\rm R}'$ the renormalization scheme in eq. \schemeone.
Let us stress that in ${\rm R}'$ the renormalized CS parameter is
equal to the bare one. Eq. \actionone\ clearly displays that the
bare parameter is shifted so that the monodromy parameter becomes
$k(1+\a).$ This is the appealing feature of ${\rm R}'$.
Notice that having a renormalized parameter equal to the bare one
is not in contradiction with renormalization theory, since CS
theory is finite. More generally, in any finite field theory
the renormalization scheme $Z_{\rm fields}=Z_{\rm parameters}=1$
is as good as any other scheme, as apposed to only renormalizable
theories, where such a scheme would not give finite renormalized
Green functions.

Another important observation [\Blasi] concerning the structure of
the radiative corrections in eq. \counter\ is that the metric only
enters in the cohomologically trivial term $\Delta X.$ This means
that changes of the metric do not reach the vacuum expectation values
of the observables and guarantees topological invariance at the
quantum level. In other words, quantum topological invariance follows
from quantum BRS invariance.

Local higher order corrections to $\gm_0$ can be constructed
recursively. Local second order radiative corrections correspond to
the solution of the equation $\Delta'\gm_2=0,$ where $\Delta'$ is the
Slavnov-Taylor operator constructed with the action $\gm'$ and
the fields $\Phi'.$ Since $\gm'=\gm_0\,[\,\Phi'\,,\,k+\a\,]\,,$
the operator $\Delta'$ is obtained from $\Delta$ by simply replacing
the fields $\Phi$ with their renormalized counterparts $\Phi'$
and $k$ with $k+\a$. This gives for $\Delta' \gm_2=0$ an equation
of the form \stability, whose solution has just been analyzed and
which leads to an expression for the local part of the effective
action up to second order of the type \actionone. In general, the
effective action up to order $n$ is given by
$\gm'=\gm_0\,[\,\Phi_{(n)}\,,\,k+\a_{(n)}\,]\,,$ with the
fields $\Phi_{(n)}$ related to the fields $\Phi_{(n-1)}$ in the same
way as $\Phi'$ are related to $\Phi$ in eq. \schemeone\ and with
$\a_{(n)}$ a power series in $\hbar$ going up to $\hbar^n.$
This concludes the analysis of the BRS identity.

We next study the Ward identities for the Landau gauge
supersymmetry. The absence of radiative corrections to the
ghost two-point Green function in eq. \actionone\ reveals that
$\gm'$ is not Landau supersymmetric. The question that arises then
is whether there is any field redefinition such that the effective
action in terms of the redefined fields satisfies the two Ward
identities eqs. \wardsusy\ and \wardsusybar. In what follows we
provide an answer in the affirmative to this question. Any wave
function renormalization
$$
\Phi = Z_\Phi\,\Phi'' \> ,
\eqn\schemetwo
$$
with
$$
Z_A = Z^{-1}_b = 1- {1\over 2}\a - \b \quad
Z_c Z_{\bar c} = 1 + \b - \ga \quad
Z_H Z_c^2 = 1- {1\over 2} \a - \ga \quad
Z_J = Z_{\bar c} \>,
$$
leads to the following renormalized effective action:
$$
\eqalign{
\gm'' = & \idx \biggl[\,-\,{ik\over 4\pi}\, \ee^{\m\r\n}
   \left( {1\over 2} \,A''^a_\m \partial_{\r} A''^a_\n +
           {1\over{3!}}\,f^{abc} A''^a_\m\,A''^b_\r\,A''^c_\n \right)
               - b''^a \partial A''^a \cr
& \qquad\qquad\qquad\qquad\qquad
  + (J''^{a\m} - \partial^\m\bar{c}''^a)\, D_\m^{\prime\prime ab} c''^b
        -{1\over 2}\,f^{abc}\,H''^a c''^b c''^c \biggr] \cr
& - {\a\over 2} \idx \biggl[ -\,{ik\over 4\pi} \, \ee^{\m\r\n}\,
        {1\over{3!}}\,f^{abc} A''^a_\m\,A''^b_\r\,A''^c_\n
           +f^{abc}\,(J''^{a\m}-\partial^\m\bar{c}''^a) A''^b_\m c^a \cr
& \qquad\qquad\qquad\qquad\qquad
   -{1\over 2}\,f^{abc}\,H''^a c''^b c''^c\, \biggr] \>. \cr}
\eqn\actiontwo
$$
It is straightforward to check that this action satisfies eqs.
\stora-\wardsusybar\ for the renormalized fields, thus ensuring
that $\gm''$  is both BRS invariant and Landau supersymmetric.
We will denote the renormalization scheme in eq. \schemetwo\ by
${\rm R}''.$ In this scheme the renormalized parameter is
also equal to the bare one, $k.$ Furthermore, since $\gm'$ in
eq. \actionone\ and $\gm''$ in eq. \actiontwo\ are related
by a field redefinition, the vacuum expectation values of the
observables computed from both actions (whatever they turn out
to be) are the same. Hence, the monodromy parameter in the scheme
${\rm R}''$ is also $k(1+\a).$ In this sense, the shift is still
present in the action $\gm'',$ though hidden.

Using different arguments, it has been shown [\Delduc] that the most
general solution over the space of local integrated functionals
of eqs. \stora-\gauge\ is precisely the effective action in eq.
\actiontwo\ for arbitrary $\a.$ Our analysis
then proves that the Landau gauge supersymmetry is devoid
of any meaning, since having a quantum breaking or not having it is
a matter of a field redefinition and fields are nothing but
{\it non-observable coordinates} in the functional space in which the
effective action and the Wilson loops are defined. In a different
lenguage, what makes sense are the cohomology classes defined by
$\Delta Y = 0$, with $Y$ a local integrated functional of mass
dimension three and ghost number zero. These cohomology classes are
labelled by $\a$ and each one of them contains an infinite number of
undistinguishable elements. Imposing the Landau gauge supersymmetry
at the quantum level amounts to choosing a particular representative
in a class, choice which is well known to be irrelevant.

{\bf\chapter{BRS-invariant regularization and broken Landau
supersymmetry}}

In this section we use a BRS invariant regularization method to
explicitly illustrate at first order in perturbation theory what we
have discussed at all orders in the previous section.

The need for a regularization method comes from the fact that,
although CS theory is known to be UV finite, the theory
is only renormalizable by power counting. This means that to compute
Green functions order by order in perturbation theory, a regularization
prescription must be introduced. The regularization method we will
use here consists in defining CS theory as the large mass limit of
topologically massive Yang-Mills (TMYM) theory, whose action in the
Landau gauge has the form [\Schonfeld,\Pisrao]
$$
S_m = S + S_{Y\!M} \qquad\quad
S_{Y\!M}={k\over 16\pi m}\idx F^a_{\mn}F^{a\,\mn} \>,
\eqn\maction
$$
with $S$ the CS action as given in eq. \starting, $F^a_{\mn}$
the field strength of the gauge connection $\A$ and $m$ a mass
parameter to be sent to infinity at the end of the calculations. We
will take $k>0$ so that the factor $e^{-S_m}$
ensures formal convergence of the path integral.
The theory defined by $S_m$ has a finite number of superficially
divergent 1PI Feynman diagrams so the adding of a Yang-Mills
term $S_{Y\!M}$ to the action $S$ does not completely
regularize CS theory. To take care of the
residual divergences we use dimensionally regularization. Our
method can then be viewed as a hybrid regularization that combines
a higher covariant derivative Yang-Mills term and dimensional
regularization. Let us be more precise and spend a few words
on the regularized theory.

We would first like to recall that there is a well known and
consistent prescription to deal with the Levi-Civita tensor in
dimensional regularization, namely the original prescription
of~'t Hooft and Veltman [\Veltman,\Maison]. Calculations certainly
get complicated, since evanescent operators enter in the game,
but algebraic consistency (something indispensable in any
regularization method [\Bonneau]) is ensured. The prescription
defines the $D\!$-dimensional analogue of $\ee_{\m\n\r}$ as a
completely antisymmetric object in its indices which satisfies
the properties
$$
\ee_{\m_1\m_2\m_3}\ee_{\n_1\n_2\n_3}=
   \sum_{\pi\in {\cal S}_3} \, {\rm sign}(\pi)\prod_{i=1}^3 \,
         \tilde{g}_{\m_i\n_{\pi (i)}} \qquad\quad
\ee_{\m_1\m_2\m_3}\hat{g}^{\m_3\m_4}=0 \> .
\eqn\epsdef
$$
Here $g_{\mn}=\tilde{g}_{\mn}\oplus\hat{g}_{\mn}$ is the
euclidean metric in $D$ dimensions and $\tilde{g}_{\mn}$ and
$\hat{g}_{\mn}$ its three- and $(D-3)\!$-dimensional projections
respectively, so that $\tilde g_{\mn} \tilde g^{\mn}=3$ and
$\hat g_{\mn} \hat g^{\mn}=D-3.$ Any $D\!$-dimensional vector $u^\m$
can be written as $u^\m = \tilde{u}^\m \oplus \hat{u}^\m,$ where
$\hat u^\m=\hat g^{\mn}u_\n$ and $\tilde u^\m=\tilde{g}^{\mn}u_\n.$
Objects with a hat vanish for $D=3$ and are called evanescent.
We stress that this prescription for $\ee_{\m\n\r}$ in $D$
dimensions is the only known one algebraically consistent; it has
proved successful in perturbative computations in a variety of
models, including WZW models [\BosWZW] and non-linear sigma models
[\Osborn].

Armed with this prescription, it is easy to construct a
dimensionally regularized TMYM theory that manifestly preserves
BRS invariance. One first extends the three-dimensional action $S_m$
in eq. \maction\ to $D$ dimensions, with $D$ an integer. Next, one
obtains the corresponding $D\!$-dimensional Feynman rules. Finally,
one promotes $D$ to a complex variable and defines every
$D\!$-dimensional Feynman integral entering in a Feynman diagram
using the dimensional regularization techniques in Ref. [\Collins].
We must emphasize at this point that only the algebraic properties of
the objects $\ee_{\m\r\n},\,\,{\hat g}_{\mn},\,\, {\tilde g}_{\mn}$
and $u_\m$ are retained for complex values of $D$ [\Maison].
Notice also that invariance of the $D\!$-dimensional action under
$D\!$-dimensional BRS transformations, together with the properties
of dimensionally regularized integrals, ensures that the formal BRS
identities hold for the regularized theory. The latter is the
same as saying that TMYM theory dimensionally regularized in this
way is manifestly BRS invariant.

Our regularization method thus defines CS theory as the limit
$m\to \infty$ of the limit $D\to 3$ of dimensionally regularized
TMYM theory. It is easy to realize that these two limits do not
commute and that they must be taken in this order if one wants to
define a sensible regularization. Notice that a necessary condition
to be able to take the limit $m\to \infty$ is that the limit
$D\to 3$ be finite. If singularitieas appear as $D$ goes 3, it
does not make sense to take $m \to \infty.$ It happens that the
limit $D \to 3$ is free of singularities to all orders in
perturbation theory [\gmr]. This does not only permit to
take the limit $m\to\infty$ but also proves that TMYM theory is
finite.

We have anticipated that the definition in eqs. \epsdef\ for the
$D\!$-dimensional $\ee_{\m\n\r}$ introduces evanescent operators.
Let us be more explicit about this. The problem is that the
definition in eqs. \epsdef\ makes the formal regularized theory
invariant under $SO(3)\otimes SO(D-3),$ rather than under $SO(D).$
As a result, the free gauge field propagator involves hatted and
twiddled objects in a non-trivial way. To see this, we write the
gauge field free propagator $D_{\mn}(\tp,\hp)$ in full detail
(see Ref. [\gmr] for the Feynman rules):
$$
D_{\mn}(\tp,\hp) = \dd_{\mn}(p) +  R_{\mn}(\tp,\hp) \> ,
\eqn\mastereq
$$
where for simplicity we have dropped colour indices and where
$\dd_{\mn}(p)$ and $R_{\mn}(\tp,\hp)$are given by
$$
\dd_{\mn}(p) = {4\pi\over k}\>{m \over p^2\,(p^2+m^2) }\,\,
   \left( m\, \ee_{\m\r\n}\,p^\r
          + p^2 g_{\mn} - p_\m p_\n \right)
\eqn\gaugeprop
$$
$$
\eqalign{
R_{\mn}(\tp,\hp) = {4\pi\over k}\> {m^3 \over p^2\,[(p^2)^2+m^2\,\tp^2]}
    \, \biggl[ \,\,& {\hp^2 \over p^2+m^2}\,
          \left( m\,\ee_{\m\r\n}\,p^\r + p^2 g_{\mn} +
                {m^2 \over p^2}\, p_\m p_\n \right) \cr
       & +  \tp^2 \hat g_{\mn} + \hp_\m \hp_\n -
              p_\m\hp_\n - \hp_\m p_\n  \,\, \biggr] ~. \cr }
$$
It is obvious that hatted quantities do not contribute at the tree
level, since they vanish at $D=3.$ This does not imply, however, that
they do not contribute at higher orders in perturbation theory, for
integration over the internal momenta of a Feynman diagram is prior
to taking the limit $D\to 3$ and integration may give rise to poles
in $D-3.$ Here we limit ourselves to showing that the hatted or
evanescent piece $R_{\mn}(\tp,\hp)$ does not contribute to the
limit $D\to 3$ of the one-loop diagrams we will compute (see Fig.
2). To this end, we recall [\Collins] that if the integral of
an evanescent quantity is convergent by power counting, then its
dimensionally regularized integral vanishes as $D$ approaches the
dimensionality of interest, three in our case. Accordingly, it is
enough to check that evanescent integrals arising from the
diagrams we are interested in are finite by power
counting at $D=3.$ But the latter follows straightforwardly if
one takes into account that the UV degree of $R_{\mn}(\tp,\hp)$
is $-4.$ (For a proof to all orders in perturbation theory of
the no-contribution of $R_{\mn}(\tp,\hp)$ to the limit $D\to 3$
of any Green function, see Ref. [\gmr]). We can then use
$\dd_{\mn}(p)$ as the gauge field free propagator in our
calculations. This ``effective'' propagator could have been
derived from the three-dimensional one by promoting the
three-momemtun to $D$ dimensions. Despite how appealing this
shortcut might look, one has to follow the long road we have
followed here if one wants to make sure that the evanescent
objects ensuring BRS invariance at the regularized level do
not contribute as $D$ goes to 3.

The one-loop corrections to the vacuum polarization tensor
$\Pi^{ab}_{\mn}(p)$, to the ghost self-energy $\Pi^{ab}(p)$ and
to the three-vertex $\gm^{abc}_{\m\n\r}$ computed with this
regularization prescription are [\gmr]
$$
\Pi_{\mn}^{ab}(p) = {7\over 3}\> {\cv\over 4\pi} \> \d^{ab}\,
                    \ee_{\m\r\n} \, p^\r  \qquad
\Pi(p)^{ab} = {2\over 3} \> {\cv\over k} \>\d^{ab}\,p^2 \qquad
\Gamma_{\m\n\r}^{abc} = 3\> {\cv\over 4\pi} \> f^{abc} \ee_{\m\n\r}
\eqn\selfen
$$
(plus contributions that vanish as $D$ approaches 3 and $m$ goes to
infinity). Eqs. \selfen\ give for the parameters $\a,\,\,\b$ and
$\ga$ of the previous section the following values:
$$
\a = {\cv\over k}\qquad \b = {2\over3}\>{\cv\over k} \qquad \ga = 0\>.
$$
We thus see that our regularization prescription gives for the shift
of the CS bare parameter $k$ the following one-loop result:
$$
k \to k +\cv \>.
\eqn\shifto
$$
This value for the one-loop shift of the bare CS parameter $k$ has
also been obtained using other regularization methods
[\Witten,\AGaume,\Asorey] and is in accordance with results from
canonical quantization [\Witten,\Bos].

Whereas our regularization method manifestly preserves BRS
invariance, it explicitly breaks the Landau gauge supersymmetry
of eqs. \susy\ and \susybar. To see the latter, we first have to
extend the transformations \susy\ and \susybar\ to $D$ dimensions
and then ckeck if they leave invariant the regularized action.
The extension of the transformations $\vm$ and $\bvm$ to $D$
dimensions is trivially achieved by using the $D\!$-dimensional
$\ee_{\m\n\r}$ defined earlier and by regarding all functions
and fields as defined on $\RR^D.$ It is then
very easy to see that the gauge-fixed CS action
in $D$ dimensions is invariant under the $D\!$-dimensional $\vm$
and $\bvm$ but that $S_{Y\!M}$ is not (see below). Hence, the
regularized theory is not Landau supersymmetric. The question that
then arises is whether the breaking remains after the regulator is
removed. We next show that is indeed the case.

Consider a generic function $F(\Phi)$ of the fields
$\Phi=\{ \A,\,b^a,\,c^a,\,\bar c^a \}.$ Under an infinitesimal
transformation $\Phi\to \Phi+\d\Phi$ of jacobian equal to one, the
following identity holds in the euclidean formalism: 
$$
\VEV{ \left(
{\partial F(\Phi) \over \partial\Phi}
-{\d S_{m}[\Phi] \over \d\Phi}\,F(\Phi) \right) \d\Phi }=0 \>.
\eqn\WI
$$
For $F(\Phi)=\A (x)\, \bar c^b(y)$ and the transformations in eq.
\susy, eq. \WI\ reads
$$
\VEV{\A(x) \> A^b_\n(y)} = {4\pi i\over k}\>\ee_{\m\r\n}
      \VEV{\partial_x^\r c^a(x) \> \bar c^b(y)}
+ \VEV{\A(x) \> \bar c^b(y) \> v_\n S_{Y\!M} } \>,
\eqn\WIAC
$$
where we have used that $v_\n S=0.$ In the following we explicitly
check up to first order in perturbation theory that the identity
\WIAC\ holds in the limit $D\to 3,\,\, m\to\infty.$ It will appear
that the second term on the RHS gives non-vanishing quantum
corrections without which the identity is not satisfied, thus showing
that the supersymmetry remains broken after the regulating parameters
are removed.

We start by computing $v_\n S_{Y\!M}$. After some algebra we obtain
that
$$
v_\n S_{Y\!M} = \OO_\n^{(0)} + \OO_\n^{(1)} + \OO_\n^{(2)} \>,
\eqn\svariation
$$
with
$$
\eqalign{
\OO_\n^{(0)} =& - {i\over m} \iddx \>
\ee_{\n\m\r}(\partial^\r c^a) \> \partial\partial \, A^{a\m} \cr
\OO_\n^{(1)} =& -{i\over m}\,\, f^{abc} \> \iddx \>
    \ee_{\n\m\r} (\partial^\r c^a) \left[
       ( \partial A^b) A^{c\m} + 2\, A^b_\s (\partial^\s  A^{c\m})
         - A_\s^b (\partial^\m A^{c\s})
\right] \cr
\OO_\n^{(2)} =& \> {i \over m}\,\, f^{abc} f^{bde} \> \iddx \>
\ee_{\n\m\r}(\partial^\r c^a)\, A^c_\s A^{d\s} A^{e\m}  \>. \cr
}
$$
The operators $\OO_\n^{(0)},$ $\OO_\n^{(1)}$ and $\OO_\n^{(2)}$ have
in momentum space the Feynman rules listed in Fig. 1. Calling
$G_\mn(p)$ and $G(p)$ to the two-point Green functions of the gauge
and ghost fields, the identity in eq. \WIAC\ can be recast in
momentum space as
$$
G_\mn(p)={4\pi \over k}\>\ee_{\m\r\n}p^{\r}\,G(p)
+G_{\m\r}(p) \, \Omega^\r_{~\n} (p) \,G(p) \>,
\eqn\WIACM
$$
where $\Omega^\r_{~\n} (p)$ is the 1PI Green function associated
to the second term on the RHS in eq. \WIAC. From a loop-wise
expansion we have:
$$
\eqalign{
G_{\mu\nu}(p)=&
\> \dd_{\m\n}(p)+\dd_{\m\s}(p)\, \Pi^{\s\r}(p)\, \dd_{\r\n}(p)
+O\,(1/k^3)  \cr
G(p)= & \> \dd(p)+\dd(p)\, \Pi(p)\, \dd(p) + O\,(1/k^2)  \cr
\Omega_{\mu\nu}(p)= & \> \Omega_{\mu\nu}^{(0)}(p)+
\Omega_{\mn}^{(1)}(p) + O\,(1/k^2) \> , \cr}
\eqn\expansion
$$
with $\dd_{\mn}(p)$ as in eq. \gaugeprop, $\dd(p)=\!-1/p^2$ the
ghost free propagator and $\Pi^{\s\r}(p)$ and $\Pi(p)$ given
in eq. \selfen. Inserting eqs. \expansion\ in eq. \WIACM\ and
identifying coefficients in $1/k,$ we obtain
$$
\dd_{\mn}(p)= {4\pi \over k}\>\ee_{\m\r\n}p^\r \,\dd(p)
   +\dd_{\m\r}(p) \,\, \Omega_{~~~~\nu}^{(0)\r}(p)\,\, \dd(p)
\eqn\zeroloops
$$
to order one (tree level), and
$$
\eqalign{
\dd_{\m\s}(p)\,\, \Pi^{\s\r}(p)\,\,\dd_{\rho\nu}(p) & =
   {4\pi \over k}\>\ee_{\m\r\n} p^\r \,\,\dd (p)\,\,\Pi(p)\,\,\dd(p)
 + \dd_{\m\s}(p)\,\,\Omega^{(1)\s}_{~~~~\nu} (p)\,\,\dd(p)  \cr
&+ \dd_{\m\s}(p)\,\, \Pi^{\s\r}(p)\,\, \dd^{\r\ga}(p) \,\,
              \Omega^{(0)}_{\ga\n} (p)\,\, \dd(p) \cr
& + \dd_{\m\s}(p)\,\, \Omega^{(0)\s}_{~~~~\nu} (p)\,\, \dd(p)
   \,\, \Pi(p) \,\, \dd(p)  ~ .\cr}
\eqn\oneloop
$$
to order two (one loop). The identity in eq. \zeroloops\
relates the tree-level
gauge field and ghost propagators at finite $m.$ {}From the Feynman
rules in Fig. 1 it follows that in the limit $m\to\infty$ the second
term on the RHS vanishes, whereas the first one reproduces the CS gauge
field free propagator. Showing that eq. \oneloop\ is indeed satisfied
requires more discussion. The explicit expressions of $\Pi_{\mn}(p)$
and $\Pi(p)$ in eqs. \selfen, together with the Feynman rules in Fig.
1, imply that the third and fourth terms on the RHS are finite and
of order $1/m$ so that they vanish when $D\to 3,\,\, m\to\infty.$ Eq.
\oneloop\ thus reduces in the limit $D\to 3,\,\, m\to\infty$ to
$$
\dd_{\m\s}(p)\,\, \Pi^{\s\r}(p)\,\, \dd_{\rho\nu}(p)=
{4\pi \over k}\>\ee_{\m\r\n} p^\r\,\,\dd (p)\,\,\Pi(p)\,\,\dd(p)
+ \dd_{\m\s}(p)\,\,\Omega^{(1)\s}_{~~~~\nu} (p)\,\, \dd(p) \>.
\eqn\modified
$$
In this equation everything is known except for
$\Omega^{(1)}_{\mn}(p),$ whose limit $D\to 3,\,\, m\to\infty$
we next compute.

There are five Feynman diagrams that contribute to
$\Omega^{(1)}_{\mn}(p)$ (see Fig. 2). All Feynman integrals arising
from these graphs are of the form
$$
I(p,m) = \int d^D\!q \,\,
    {{m^r \,M(q)} \over {\prod_{i} (l_i^2 +m_i^2)^{s_i}}}
            \quad  \qquad r,s_i \in {\rm I\!\!\, N} \>,
$$
where $M(q)$ is a monomial of degree $n_q$ in the components of the
integrated momentum $q,$ the vectors $l_i$ are linear combinations
of $q$ and the external momenta $p_1,\ldots ,p_E,$ and the masses
only take on two values, $m_i=0$  and $m_i=m>0.$ The external momenta
are assumed to lie in a bounded subdomain of ${\RR}^{D}.$ As we
have already said, we first have to take the limit $D\to 3$
of $I(p,m)$ and then $m\to \infty.$ The limit $D\to 3$ is always
finite, for in dimensional regularization the integral $I(p,m)$ is
finite as $D$ approaches $l$ for $l$ odd, even when $I(p,m)$ is
divergent by power counting [\Speer]. This guarantees that no poles
appear when the limit $D\to 3$ is taken. To compute the large $m$
limit of $I(p,m)$ at $D=3,$ hence of the diagrams we are interested
in, we use two vanishing theorems. Here we limit ourselves to state
them. Their proof and generalization to higher orders in
perturbation theory can be found in Ref. [\gmr]. Denoting by $d$ the
mass dimension of $I(p,m)$ and introducing the notation $[n]=0$ for
$n$ even and $[n]=1$ for $n$ odd, the theorems say that

{
\leftskip=0.8 true cm \rightskip=0.8 true cm
\noindent {\bf Theorem 1:} {\sl If $I(p,m)$ is infrared convergent
by power counting, $ d < 0 $ and $\a m-2\sum_i \b_i m_i \, < \,0,$
then $I(p,m)$ vanishes when $m$ goes to $\infty.$}
\par
}

{
\leftskip=0.8 true cm \rightskip=0.8 true cm
\noindent {\bf Theorem 2:} {\sl If $I(p,m)$ is absolutely convergent
by power counting for exceptional configurations of the external
momenta and $ [n_q] > d,$ then $I(p,m)\to 0$ as $m\to \infty.$}
\par
}

\noindent After taking the limits $D\to 3,\,\, m\to\infty$ and using
the theorems, we obtain for the diagrams in Fig. 2 the following
results:\foot{The algebra was performed with the help of the symbolic
language REDUCE [\reduce].}
$$
\eqalign{
\D_1 = & \, {\cv\over k}\> \left( - {13\over 15}\> p^2\,\d_{\mn}
         + {19\over 15}\> p_\m p_\n
         + m\,\ee_{\mn\r}\, p^\r \right) \cropen{11pt}
\D_2 = & \, {\cv\over k}\> \left( {16 \over 15}\>p^2\,\d_{\mn}
         - {18 \over 15}\> p_\m p_\n
         - {2\over 3}\> m\,\ee_{\mn\r}\, p^\r \right) \cropen{11pt}
\D_3 = & \, {\cv\over k}\> \left( {8\over 3}\>p^2\,\d_{\mn}
         - {8\over 3}\> p_\m p_\n
         - 2\,m\,\ee_{\mn\r}\, p^\r \right) \cropen{11pt}
\D_4 = & \, {\cv\over k}\> {4\over 3}\> m\,\ee_{\mn\r}\, p^\r \cropen{11pt}
\D_5 = & \, {\cv\over k}\> \left( {2\over 15}\>p^2\,\d_{\mn}
         - {6\over 15}\> p_\m p_\n
         + {1\over 3}\>m\,\ee_{\mn\r}\, p^\r \right) \cr}
$$
Any other contribution vanishes as $D$ goes to 3 and $m$ approaches
infinity. Summing over diagrams we finally have:
$$
\Omega_{\s\n}^{(1)}(p) = 3\> {\cv\over k} \>
             (p^2\> \delta_{\s\n}  - p_\s p_\n) \> .
\eqn\breaking
$$
Note that contributions of order $m$ from individual diagrams cancel
when summing over diagrams, thus making the limit $m\to\infty$
well defined. From eqs. \selfen\ and \breaking\ it follows that the
identity \modified\ is verified. It is very important to
realize that were it not for the non-vanishing contribution
$\Omega_{\s\n}^{(1)}(p),$ the identity \modified\ would not hold.
Recalling that $\Omega_{\s\n}^{(1)}(p)$ had its origin in the
supersymmetry breaking term in the regularized action, we
conclude that the supersymmetry remains broken after the regulating
parameters are removed and that it is precisley the breaking what is
required to have the identity \modified\ satisfied. This is not
peculiar of the regularization method used here but has also been
observed [\Dorey] for a hybrid regulator consisting of a higher
covariant derivative term of the form $(DF)^2$ and Pauli-Villars
[\AGaume].

The same pattern occurs for the Landau gauge supersymmetry in eq.
\susybar. If in eq. \WI\ we take $F(\Phi)=\A(x)\,c^b(y)$ and
the transformation $\bvm$ in eqs. \susybar, we get the identity
$$
\VEV{\A(x) \> A^b_\n(y)} = {4\pi i\over k}\>\ee_{\m\r\n}
      \VEV{\partial_x^\r c^a(x) \> \bar c^b(y)}
     + \VEV{c^a(x)\, A^b_\n(y)\, \bvm S_{Y\!M} } \>.
\eqn\WIABC
$$
This identity can be analyzed in exactly the same way as the one
in eq. \WIAC. As a matter of fact, both identities have the same
form in momentum space, namely eq. \WIACM. One can think of the
identity eq. \modified\ as a consistency check for the
one-loop corrections to the vacuum polarization tensor and to the
ghost self-energy in eq. \selfen. In a similar way one can check
the value for the one-loop correction $\gm^{abc}_{\m\n\r}$ to the
three-vertex. In this case, it is enough to take
$F(\Phi)=\A(x)\, A^b_\n (y)\, \bar c^c(z)$ and the transformation
$v_\r$ in eq. \susy.

{\bf\chapter{Conclusions}}

We have explicitly shown in Sect. 2 that the Landau gauge
supersymmetry of CS theory [\Birmingham,\Gieres,\Delduc] does
not have any significance. We
have done this by proving that having a quantum breaking of the
supersymmetry or not having it is only a question of a wave
function renormalization which does not affect the vacuum
expectation values of the observables. Morever, we have given
two expressions $\gm'$ and $\gm''$ for the local part of the
renormalized effective action, both yielding the same
vacuum expectation values of the Wilson loops (whatever those
turn out to be), but one of them $(\gm'')$ being Landau
supersymmetric and the other one $(\gm')$ not.

This observation, combined with the fact that topological
invariance is recovered from BRS invariance
and the prediction of a shift on the grounds
of only BRS invariance [see eg. eq.\effective], leaves us with
BRS as the only fundamental symmetry of the theory.

To compute the actual value of the shift of the bare CS parameter,
a regularization prescription is needed if one insists in employing
Feynman diagrams. Using a regularization prescription manisfestly
preserving the Landau supersymmetry is of no importance, as
far as it preserves BRS invariance. The reason is that any BRS
invariant regularization prescription will yield an action of the
form in eq. \effective, which can always be recast in the Landau
supersymmetric fashion \actiontwo, both actions being physically
undistinguishable.

In Sect. 3 we have provided an example of a BRS invariant
regularization method that breaks the Landau gauge supersymmetry
and have checked that the latter supersymmetry remains broken
after the regulating parameters are removed. The shift of the bare
CS parameter as computed with this method is $k\to k+ \cv,$
in agreement with results from canonical quantization. In the
following table we collect in units of $\cv/k$ the one-loop results
for $\a,\,\,\b$ and $\ga$ in eq. \effective\ as computed with all
BRS invariant regulators tried so far in CS theory:\foot{The
values given here for higher covariant derivatives plus
Pauli-Villars are those computed in Ref. [\review] rather than
those in Ref. [\AGaume], where strictly speaking only Pauli-Villars
fields and no higher covariant derivative terms are used.
\nextline
$\qquad$ Geometric regularization makes use of ghost generations
different from the standard Fad\-deev-Popov ones, so only the pure
gauge sector of the renormalized effective action can be compared.
The quantity $I_n$ is defined as $$I_n=\int_0^\infty dp\>\>
{(1+p^2)^n \over 1+p^2(1+p^2)^{2n} }\>,$$ with $n>1$ an integer.}
$$
\vbox{
\offinterlineskip
\def\tablerule{\noalign{\hrule}}
\def\homit{height3pt&\omit&&\omit&&\omit&&\omit&\cr}
\hrule
\halign{&\vrule#&
\strut\quad\hfil#\hfil\quad\cr
\homit
& $\qquad\>$ Regularization Method \hfil&& $\a$ && $\b$ && $\ga$ &\cr
\homit
\tablerule
\homit
& Method in Sect. 3 && 1 && 2/3 && 0 &\cr
\tablerule
\homit
& $\eta$-function regularization [\Witten] && 1 && 0 && 0 &\cr
\tablerule
\homit
&Higher covariant derivatives + Pauli-Villars [\review]
                                          && 1 && 2/9 && 0 &\cr
\tablerule
\homit
& Geometric regularization [\Asorey]
                         && 1 && $4I_n/3\pi$ && $-$ &\cr
\homit }
\hrule
}
$$

As can be seen, different BRS invariant regularization methods
give different values for $\b$ and $\ga$ but the same value for the
shift $\a.$ This uniqueness for the value of $\a$ for all BRS
invariant regulators tried as yet suggests parametrizing the quantum
theory in terms of the bare parameter:
$$
k_{\rm renormalized} = k_{\rm bare} = k \>.
$$
The idea behind this parametrization is that the quantum theory is
unambiguously construct-ed by BRS invariance, if preserved at the
regularized level. Notice that such a pa\-ra\-me\-tri\-za\-tion would be
nonsensical if two different BRS preserving regulators yielded
different values for $\a,$ but the results in the table show that
for all BRS invariant regulators tried to date this is not the case.
CS theory thus gives a concrete realization of the idea that, in
a finite theory, the bare parameters constitute the right
parametrization of the quantum theory, provided one uses
regulators preserving the fundamental symmetries of the theory
[\Jaffe].

The agreement on the value of $\a$ for different BRS invariant
regulators can not be explained within the framework of local
perturbative renormalization theory [\Epstein], for, according
to its principles, the ambiguities introduced by any regularization
method should reach the value of $\a.$ Note also that local
perturbative renormalization theory does not contemplate the
idea of a preferred parametrization. Any argument aiming to
choosing a particular parametrization has to be found outside this
framework. Here we have used the argument of the symmetries
characterizing the theory.

It would be desirable to learn whether the one-loop agreement
of the table holds at higher orders. We conjecture that
this is the case. Unfortunately, no comparison is possible,
since so far only the regularization method[ proposed here has
produced a two-loop computation of the shift [\gmr], with the
result that there is no second-order correction to the one-loop
result, in agreement with canonical quantization.

\bigskip

\ack

\noindent
The authors are grateful to G. Bonneau, F. Delduc and  G. Leibbrandt
for valuable conversations. GG was supported by The Commission of the
European Communities through contract No. SC900376, CPM by the Natural
Sciences and Engineering Research Council of Canada under grant No.
A 8063, and FRR by The Commission of the European Communitites through
contract No. SC1000488 with The Niels Bohr Institute and by the Stichting
voor Fundamenteel Onderzoek der Materie of The Netherlands. They also
acknowledge partial support from Comisi\'on de Investigaci\'on
Cient\'\i fica y T\'ecnica, Spain.

\refout

\vfil\eject





\centerline{{\fourteenbf Figures' captions}}

\bigskip

Figure 1: Feynman rules for the the operators $\OO_\n^{(0)},\,\,
\OO_\n^{(1)}$ and $\OO_\n^{(2)}.$

\bigskip

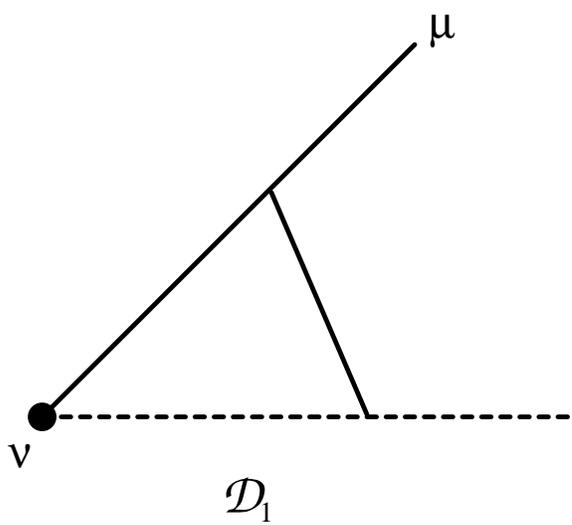
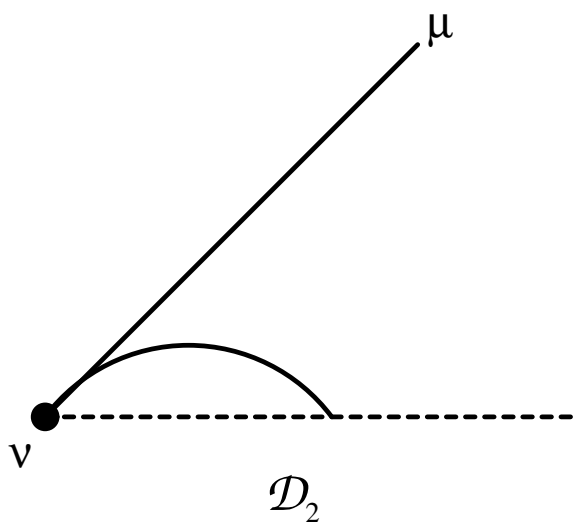
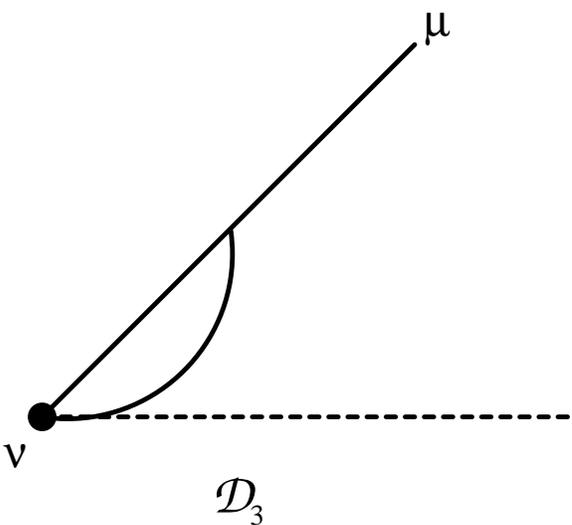
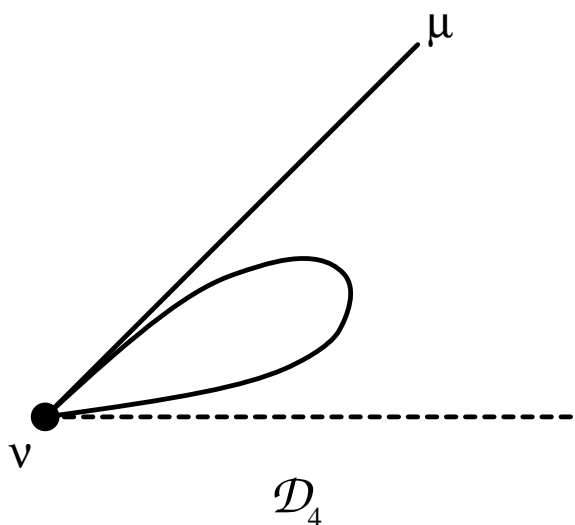
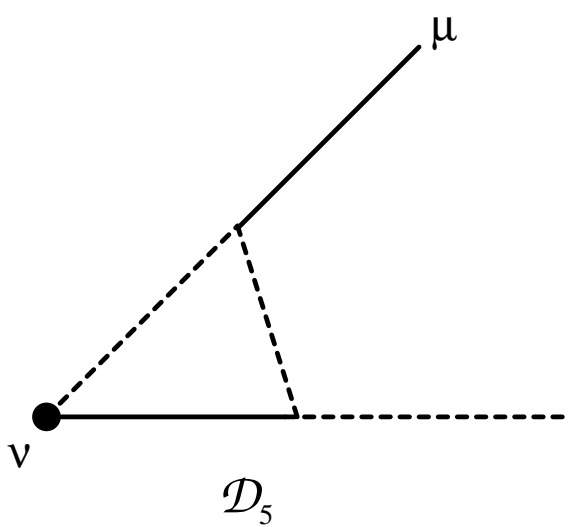
Figure 2: Feynman diagrams contributing to $\Omega_{\m\n}^{(1)}(p).$

\bye